\documentclass[prc,preprint,showpacs,preprintnumbers,amsmath,amssymb]{revtex4}

\usepackage[mathscr]{eucal}
\usepackage{graphicx}
\def\slr#1{\setbox0=\hbox{$#1$}           
   \dimen0=\wd0                                 
   \setbox1=\hbox{/} \dimen1=\wd1               
   \ifdim\dimen0>\dimen1                        
      \rlap{\hbox to \dimen0{\hfil/\hfil}}      
      #1                                        
   \else                                        
      \rlap{\hbox to \dimen1{\hfil$#1$\hfil}}   
      /                                         
   \fi}

\def\gev#1{ GeV${}^{#1}$}
\def\be{\begin{eqnarray}}
\def\ee{\end{eqnarray}}

\renewcommand{\theequation}%
    {\arabic{section}.\arabic{equation}}
\makeatletter \@addtoreset{equation}{section} \makeatother

\begin{document}

\preprint{BCCNT: 05/101/336}

\title{Vector-Isovector Excitations at Finite Density and Temperature and the Validity of Brown-Rho Scaling}

\author{Xiangdong Li}

\affiliation{%
Department of Computer System Technology\\
New York City College of Technology of the City University of New
York\\
Brooklyn, New York 11201 }%

\author{C. M. Shakin}
\email[email:]{cshakin@brooklyn.cuny.edu}

\affiliation{%
Department of Physics\\
Brooklyn College of the City University of New York\\
Brooklyn, New York 11210
}%

\date{\today}

\begin{abstract}

Recent work reported at the Quark Matter Conference 2005 has led
to the suggestion that Brown-Rho scaling is ruled out by the NA60
data. (Brown-Rho scaling describes the reduction of hadronic
masses in matter and at finite temperature.) In the present work
we argue that the interpretation of the experimental data
presented at the Quark Matter Conference is not correct and that
Brown-Rho scaling is valid. To make this argument we discuss the
evolution in time of the excited hadronic system and suggest that
the system is deconfined at the earliest times and becomes
confined when the density and temperature decrease as the system
evolves. Thus, we suggest that we see both the properties of the
deconfined and confined systems in the experimental data. In our
interpretation, Brown-Rho scaling refers to the later times of the
collision, when the system is in the confined phase.

\end{abstract}


\maketitle

\section{INTRODUCTION}

There has been an ongoing attempt to understand how quantum
chromodynamics, the theory of strong interactions, governs the
properties of hadrons in vacuum and in matter. Associated with
this program are investigations of the properties of the
quark-gluon plasma and of hadronic matter at finite temperature
and finite matter density. (Experimental data concerning mesons in
matter are often discussed in terms of Brown-Rho (BR) scaling [1].
Recent reviews may be found in Refs. [2,3].)

Various authors have discussed the behavior of meson masses in
matter. For example, Hatsuda and Lee obtain \be
m_\rho^*=m_\rho(1-0.18\,n_B/n_0),\ee using QCD sum rules [4]. Here
$n_B$ is the baryon density and $n_0$ is the density of nuclear
matter. The Brooklyn College Group has also studied the properties
of mesons at finite temperature and finite density [5-12] and we
will make use of their results as we proceed.

The experimental data [13] obtained by the NA60 experiment is
shown in Fig. 1 [14]. There the solid curve corresponds \be
m_\rho^*=m_\rho(1-0.15\,n_B/n_0),\ee while the dashed curve
represents \be
m_\rho^*=m_\rho(1-0.15\,n_B/n_0)\,\,(1-[T/T_c]^2)^{0.3}.\ee

\begin{figure}
\includegraphics[bb=0 300 450 450, angle=0, scale=1.5]{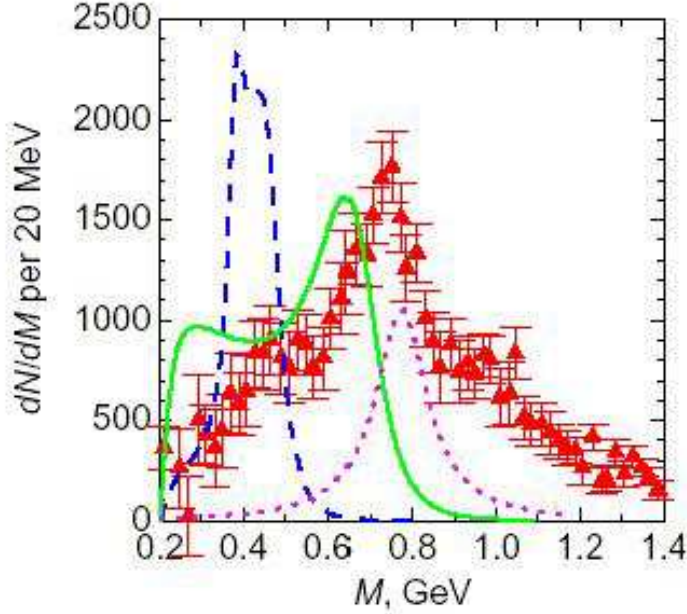}%
 \caption{Invariant mass distribution of dimuons from semi-central
In+In collisions at the beam energy $158$A GeV. Experimental
points are from [13]. The solid and dashed curves are calculated
using the $\rho$-mass modification factors of Eqs.\,(1.2) and
(1.3), respectively. The dotted line indicates the
hydrodynamically calculated $\rho$-meson decay at the freeze-out.
(This figure appears in Ref. [14].)}
\end{figure}

The dilepton rate is given in Ref. [14] as \be
\frac{d^8N}{d^4xd^4q}=-\mathcal{L}(M)\,\,\frac{\alpha^2}{\pi^3q^2}\,\,f_B(q_0,1/T)\,\,\mbox{Im}\Pi_{em}(q,T,\mu_b),\ee
where $q^2=M^2=q_0^2-\vec{q}\,\,^2$. The Bose distribution
function is \be f_B(q_0,\beta=1/T)=(e^{\beta q_0}-1)^{-1}, \ee and
the lepton kinematic factor is \be
\mathcal{L}(M)=\left(1+\frac{2m_l^2}{M^2}\right)\,\,\sqrt{1-\frac{4m_l^2}{M^2}},\ee
with lepton mass $m_l$. If the pole shift is neglected, the
imaginary part of the current correlation function is [14] \be
\mbox{Im}\Pi_{em}(M)=\frac{m_\rho^4}{g^2}\,\,\frac{\mbox{Im}\Pi(M)}{(M^2-m_\rho^2)^2+\mbox{Im}\Pi^2(M)},
\ee with \be
\mbox{Im}\Pi(M)=-\frac{g^2_{\rho\pi\pi}}{48\pi}\,\,\frac{(M^2-4m^2_\pi)^{3/2}}{M},\ee
\be \approx -\frac{g^2_{\rho\pi\pi}}{48\pi}M^2.\ee In the present
work we will present results for the hadronic correlation function
obtained in earlier studies for values of $T>T_c$ and $T<T_c$
[5-12].

In recent years we have developed a generalized Nambu-Jona-Lasinio
(NJL) model that incorporates a covariant model of confinement.
The Lagrangian of the model is \be {\cal L}=&&\bar q(i\slr
\partial-m^0)q +\frac{G_S}{2}\sum_{i=0}^8[
(\bar q\lambda^iq)^2+(\bar qi\gamma_5 \lambda^iq)^2]\nonumber\\
&&-\frac{G_V}{2}\sum_{i=0}^8[
(\bar q\lambda^i\gamma_\mu q)^2+(\bar q\lambda^i\gamma_5 \gamma_\mu q)^2]\nonumber\\
&& +\frac{G_D}{2}\{\det[\bar q(1+\gamma_5)q]+\det[\bar
q(1-\gamma_5)q]\} \nonumber\\
&&+ {\cal L}_{conf}\, \ee where the $\lambda^i(i=0,\cdots, 8)$ are
the Gell-Mann matrices, with $\lambda^0=\sqrt{2/3}\mathbf{\,1}$,
$m^0=\mbox{diag}\,(m_u^0, m_d^0, m_s^0)$ is a matrix of current
quark masses and ${\cal L}_{conf}$ denotes our model of
confinement. Many applications have been made in the study of
light meson spectra, decay constants, and mixing angles. In the
present work we describe the use of our model when we include a
description of deconfinement at finite density and temperature.

The organization of our work is as follows. In Section II we
review results for mesonic excitations at finite matter density
and at zero temperature. (It is of interest to note that the pion
and kaon masses do not change very much with increasing density or
temperature since these mesons are (pseudo) Goldstone bosons.) In
Section III we discuss the properties of mesons at finite
temperature in the confined mode. In Section IV we consider
temperatures above $T_c$, the temperature for deconfinement. Above
$T_c$, one finds resonant structures in the plasma which
correspond to some of the excitations seen in computer simulations
of QCD [15-22] whose analysis makes use of the maximum entropy
method (MEM). Such excitations are also seen in our calculations
made at finite density and zero temperature and are described in
Section V.

In Section VI we return to a discussion of the NA60 data of Fig. 1
and argue that Brown-Rho scaling is indeed correct and that the
experimental data supports our observation that the compound
system evolves from the deconfined to the confined mode as the
collision develops in time. Section VII contains some further
comments and conclusions. Finally, in the Appendices, we review
our model of confinement at finite temperature and at finite
density.

\section{Mesonic Excitations at Finite Density}

 \begin{figure}
 \includegraphics[bb=0 15 300 220, angle=0, scale=1]{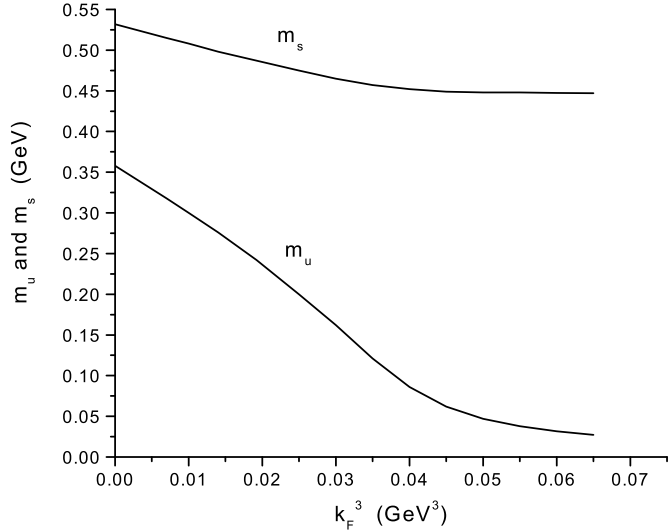}%
 \caption{The density-dependent constituent
 quark masses, $m_u(\rho)=m_d(\rho)$ and $m_s(\rho)$ are shown.
 (See Ref. [12] and caption to Fig. 3.).}
 \end{figure}

\begin{figure}
 \includegraphics[bb=0 10 300 220, angle=-0.5, scale=1]{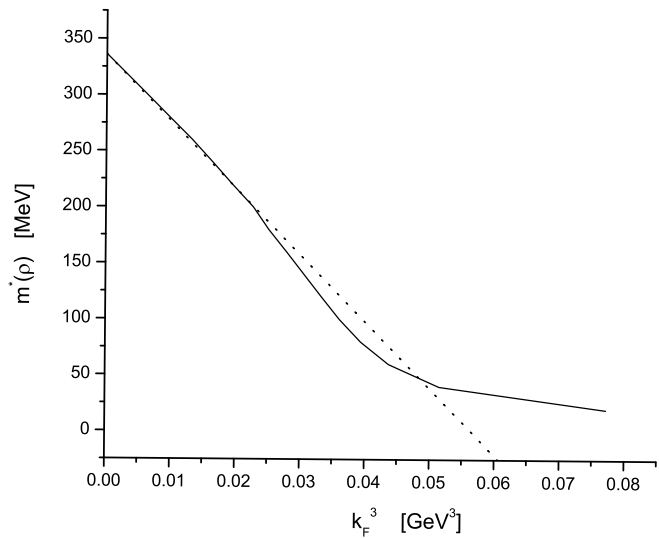}%
 \caption{The dashed line is a linear approximation to the result shown in Fig. 2 which we use for
$\rho\leq2\rho_{NM}$.  (Nuclear matter density corresponds to
$k_F^3=0.0192$ \gev3.) See Ref. [12].}
 \end{figure}

 \begin{figure}
 \includegraphics[bb=0 60 280 280, angle=0, scale=0.8]{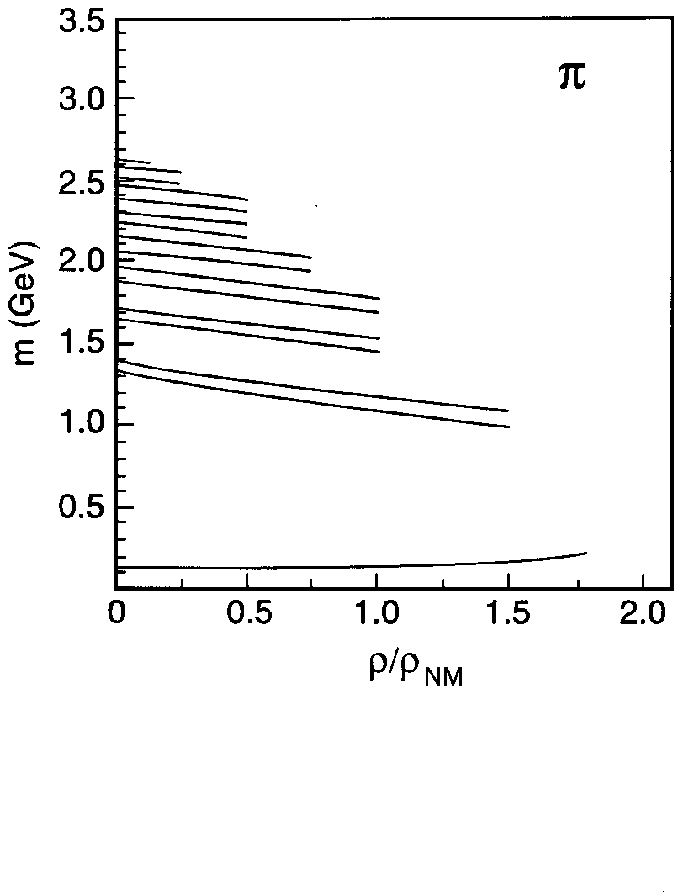}%
 \caption{The mass values for the pion and its radial excitations are presented as a function of
 the density of matter. Here, the NJL model was used with density-dependent coupling constants,
 density-dependent masses and a density-dependent confining interaction [12]. [See the
 Appendices.] We use $G_\pi(\rho)=G_\pi(0)[1-0.087\rho/\rho_{NM}]$
 and $m_u(\rho)=m_d(\rho)=m_u^0+0.3585\,\mbox{GeV}[1-0.4\rho/\rho_{NM}]$,
 with $m_u^0=0.0055$ GeV. We use $G_\pi(0)=13.49$\gev{-2} and $G_V=11.46$\gev{-2}.
 Note that the various curves end at densities beyond which the excitations are no longer bound states.}
 \end{figure}

\begin{figure}
 \includegraphics[bb=0 30 280 220, angle=0, scale=0.8]{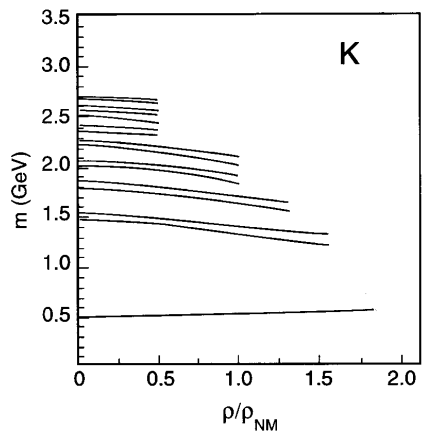}%
 \caption{Mass values of the $K$ mesons are shown as a function of the density of matter.
 Here we use $G_K(0)=13.07$\gev{-2},
 $G_K(\rho)=G_K(0)[1-0.087\rho/\rho_{NM}]$ and
 $G_V=11.46$\gev{-2}. (See Ref. [12].)}
 \end{figure}

 \begin{figure}
 \includegraphics[bb=0 90 400 400, angle=0, scale=0.6]{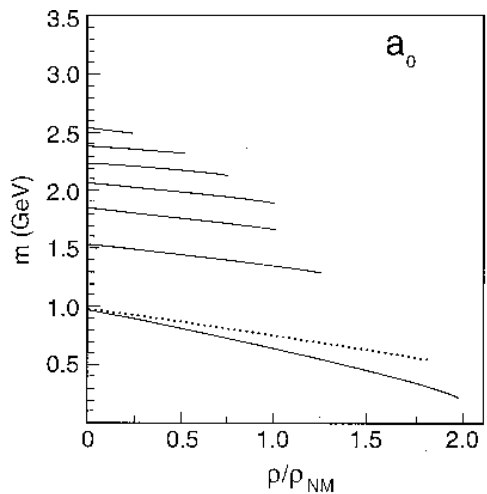}%
 \caption{Mass values for the $a_0$ mesons are given as a function of the matter density.
 Here, we have used $G_{a_0}(0)=13.10$\gev{-2} and
 $G_{a_0}(\rho)=G_{a_0}(0)[1-0.045\rho/\rho_{NM}]$. We have also
 used $m_u=m_u^0+0.3585\,\mbox{GeV}[1-0.4\rho/\rho_{NM}]$
 with $m_u^0=0.0055$ GeV. The dotted line results if we put
 $G_{a_0}(\rho)=G_{a_0}(0)[1-0.087\rho/\rho_{NM}]$ and use the mass values of
 Table I of Ref. [12]. The dotted curve is similar to the curve for the
 $a_0$ mass given in Ref. [23]. The curves representing the masses of
 the radial excitations are changed very little when we use the second
 form for $G_{a_0}(\rho)$ given above. The dotted curve is reasonably well represented by
 $m^*_{a_0}(\rho)=m_{a_0}[1-0.22\rho/\rho_{NM}]$.}
 \end{figure}

  \begin{figure}
 \includegraphics[bb=0 70 280 270, angle=0, scale=0.8]{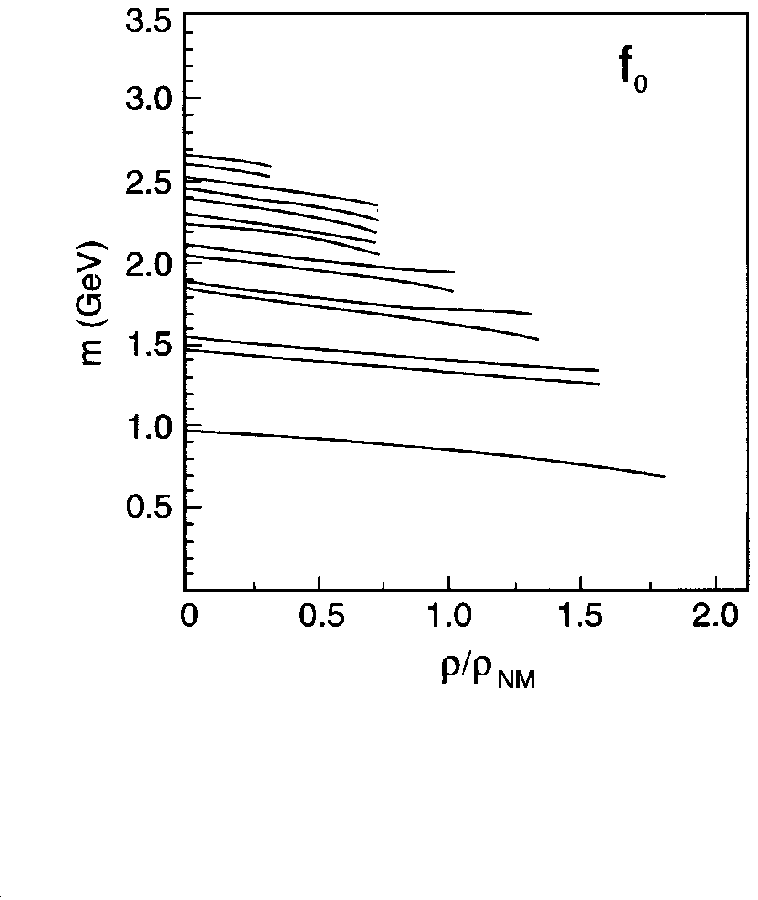}%
 \caption{The figure shows the mass values of the $f_0$ mesons as a
 function of density. The mass values for the quarks are taken from
 Table I of Ref. [12]. In a singlet-octet representation, we have used the
 constants $G_{00}^S=14.25$\gev{-2}, $G_{08}^S=0.4953$\gev{-2} and $G_{88}^S=10.65$\gev{-2}.
 Deconfinement takes place somewhat above $\rho=1.8\rho_{NM}$.
 (See Ref. [12].) For small $\rho/\rho_{NM}$, the mass of the $f_0$ is fairly well represented by
 $m^*_{f_0}(\rho)=m_{f_0}[1-0.14\rho/\rho_{NM}]$.}
 \end{figure}

  \begin{figure}
 \includegraphics[bb=0 50 280 270, angle=0, scale=0.8]{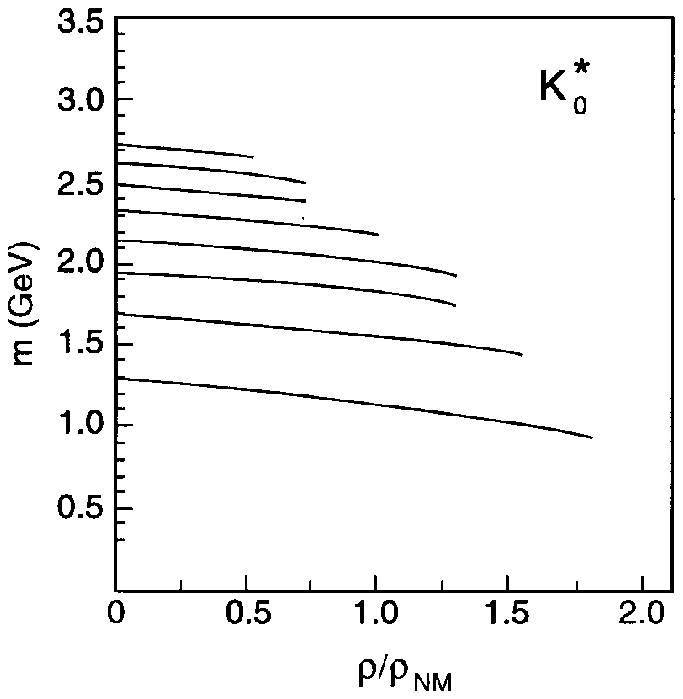}%
 \caption{The figure shows the mass values obtained for the $K_0^*$ mesons as a
 function of density. Here we use a constant $G_{K_0^*}=10.25$\gev{-2}.
 Deconfinement takes place somewhat above $\rho=1.8\rho_{NM}$. For the smaller values of
 $\rho/\rho_{NM}$ the mass of the $K_0^*$ is given by $m^*_{K_0^*}(\rho)=m_{K_0^*}[1-0.14\rho/\rho_{NM}]$.}
 \end{figure}

In this section we review some of the results reported in Ref. [5]
for meson mass values at finite matter density. We made use of the
density-dependent masses which are shown in Figs. 2 and 3. We also
used density-dependent coupling constants in a generalized NJL
model, in part to avoid pion condensation, and we have also
introduced a density-dependent confining interaction [5]. Our
model of confinement is discussed in Appendices A and B.

In Fig. 4 we see  results for the pion and its various radial
excitations. Note that the pion energy is fairly constant up to
the point of deconfinement, since the pion is a (pseudo) Goldstone
boson. In this case, deconfinement takes places at
$\rho/\rho_{NM}\simeq 1.75$. In Fig. 5 we show corresponding
results for the $K$ meson. Figs. 6 and 7 show our results for the
$a_0$ and $f_0$ mesons, respectively. (The curve for the $f_0$
meson is reasonably well fit with
$m_{f_0}^*=m_f(1-0.15\rho/\rho_{NM})$ for $\rho/\rho_{NM} <1.0$).
Finally, in Fig. 8 we show our results for the $K_0^*$ meson mass
as a function of density [12]. (In these calculations we have used
the covariant confinement model which we review in the
Appendices.)

The results presented in this Section are generally consistent
with Brown-Rho scaling at finite density as represented by Eq.
(1.1), for example. We see that our results are in agreement with
those of Hatsuda and Lee [4], although we have used an entirely
different method of calculation.

\newpage

\section{Mesonic Excitations at Finite Temperature}

 \begin{figure}
 \includegraphics[bb=0 15 300 220, angle=0, scale=0.8]{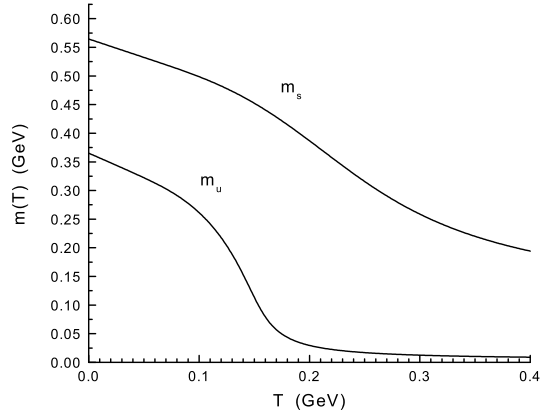}%
 \caption{The temperature-dependent constituent
 quark masses, $m_u(T)$ and $m_s(T)$, are shown. Here $m_u^0=0.0055$
 GeV, $m_s^0=0.130$ GeV, and $G_S(T)=5.691[1-0.17(T/T_c)]$, if we
 use Klevansky's notation [24]. We have used the equation
 $m(T)=m^0+2G_S(T)N_c\frac{m(T)}{\pi^2}\int_0^\Lambda
 dp\frac{p^2}{E_p}\tanh(\frac{1}{2}\beta E_p),$ which appears in
 Ref. [24].}
 \end{figure}

Calculations similar to those made at finite density have also
been made at finite temperature [11]. In this case,
temperature-dependent coupling constants were used in our
generalized NJL model. The quark masses were calculated as a
function of temperature and are shown in Fig. 9. Also, a
temperature-dependent confining interaction was used based upon
the lattice QCD analysis of Ref. [25]. (See Appendix B.)

 \begin{figure}
 \includegraphics[bb=0 15 440 670, angle=0, scale=0.3]{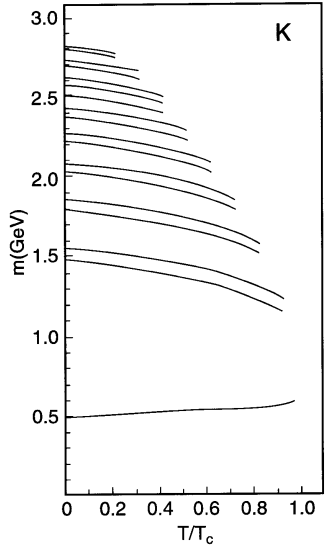}%
 \caption{Mass values of kaonic states calculated with $G_K(T)=
 13.07[1-0.17\,T/T_c]$ GeV, $G_V(T)=11.46[1-0.17\,T/T_c]$ GeV, and the quark
 mass values given in Fig. 9. The value of the kaon mass is 0.598 GeV at
 $T/T_c=0.95$, where $m_u(T)=0.075$ GeV and $m_s(T)=0.439$ GeV. (See Ref. [11].)}
 \end{figure}

 \begin{figure}
 \includegraphics[bb=0 0 440 670, angle=0, scale=0.4]{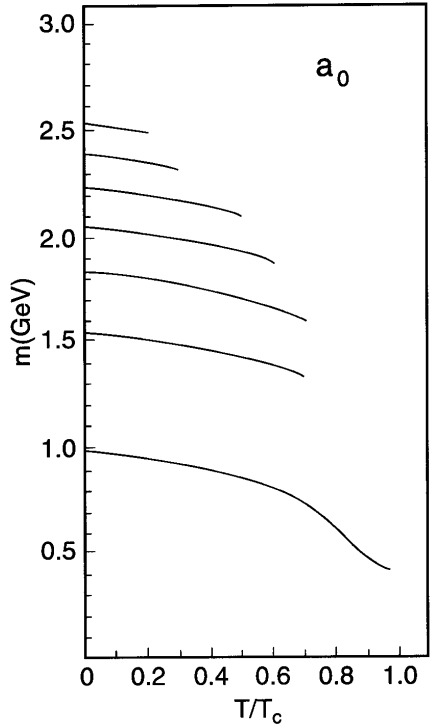}%
 \caption{Mass values for the $a_0$ mesons calculated with $G_{a_0}(T)=
 13.1[1-0.17\,T/T_c]$ GeV, and the quark mass values given in Fig. 9.
 The value of the $a_0$ mass at $T/T_c=0.95$ is 0.416 GeV. (See Ref. [11].)}
 \end{figure}

 \begin{figure}
 \includegraphics[bb=0 15 440 670, angle=0, scale=0.4]{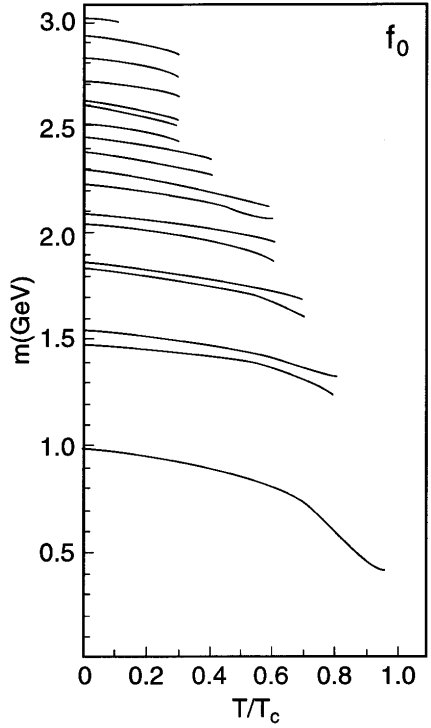}%
 \caption{Mass values of the $f_0$ mesons calculated with
 $G_{00}(T)=14.25[1-0.17\,T/T_c]$ GeV, $G_{88}(T)=10.65[1-0.17\,T/T_c]$ GeV,
 $G_{08}(T)=0.495[1-0.17\,T/T_c]$ GeV, and $G_{80}(T)=G_{08}(T)$ in a singlet-octet
 representation. The quark mass values used are shown in Fig. 9. The
 $f_0$ has a mass of 0.400 GeV at $T/T_c=0.95$. (See Ref. [11].)}
 \end{figure}

 \begin{figure}
 \includegraphics[bb=0 15 440 670, angle=0, scale=0.3]{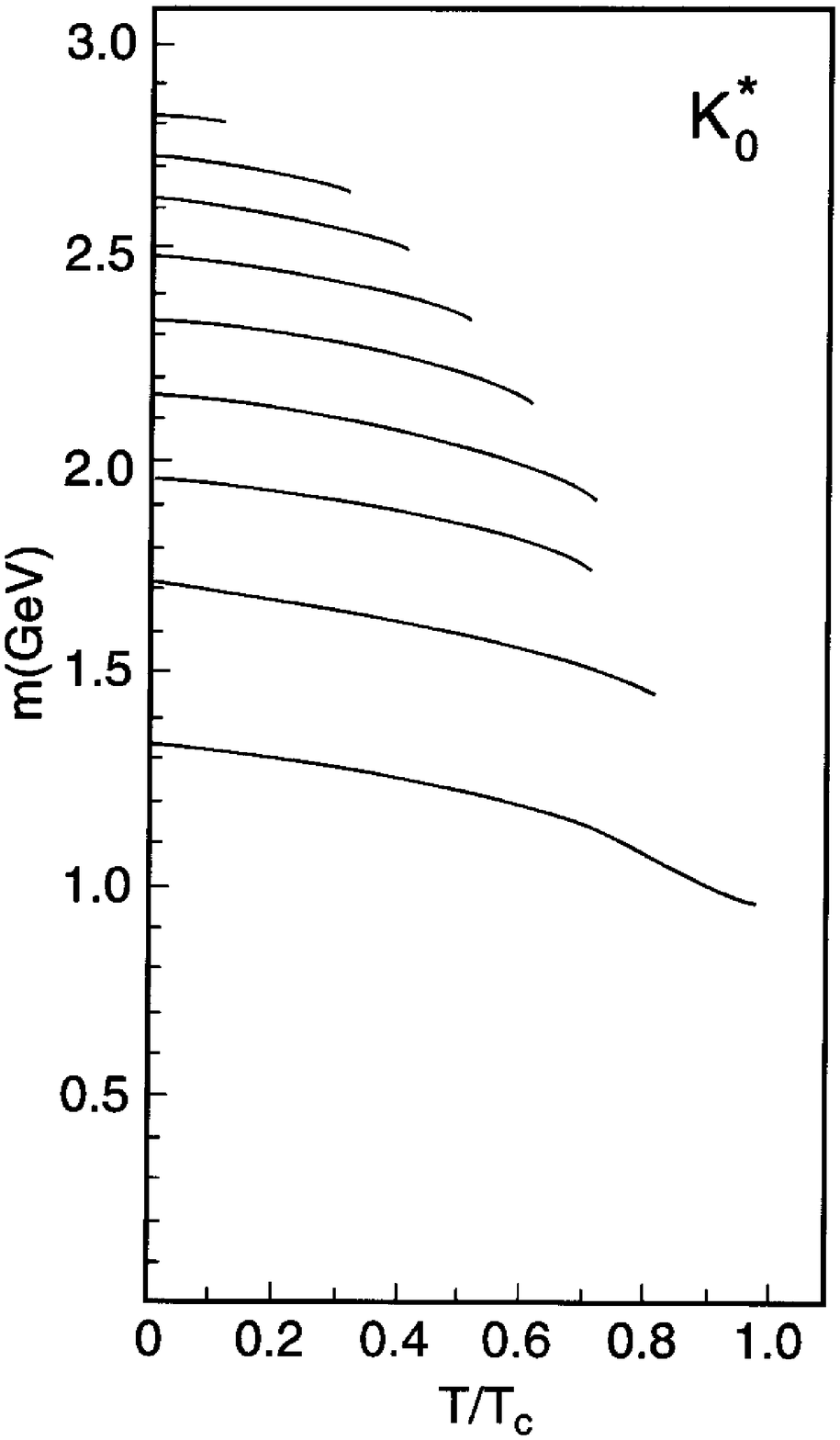}%
 \caption{Mass values obtained for the $K_0^*$ mesons calculated using $G_{K_0^*}(T)=
 10.25[1-0.17\,T/T_c]$ GeV and the quark mass values shown in Fig. 9. (See Ref. [11].)}
 \end{figure}

In Figs. 10-13 we show our results for the $K$, $a_0$, $f_0$ and
$K_0^*$ mesons. In all these cases the mesons are no longer
confined at energies slightly below $T_c$. (See Figs. 10-13.) In
these calculations we find that the linear approximation
$m^*(T)=m_0[1-\alpha T/T_c]$ is only satisfactory up to about
$T/T_c\simeq0.5$ for the $a_0$, $f_0$ and $K_0^*$ mesons.

\newpage

\section{Excitations of the Quark-Gluon Plasma at finite Density}

For calculations at finite density we use the formalism of Ref.
[8]. In this case the unusual form of the curves shown in Figs. 14
and 15 is due to Pauli blocking of the excitations by the filled
states of the Fermi sea of quarks at finite density and zero
temperature.

\begin{figure}
\includegraphics[bb=0 0 300 210, angle=0, scale=1.2]{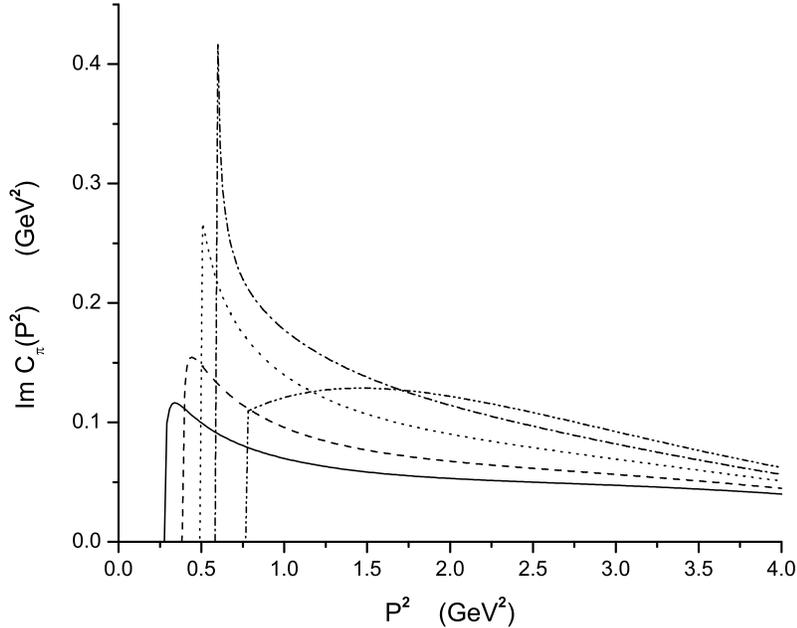}%
\caption{The figure presents values of the correlation function,
$\mbox{Im}C_\pi(P^2)$, for various values of $\rho/\rho_c$. Here,
$\rho/\rho_c=1.2$ [solid line], 2.0 [dashed line], 3.0 [dotted
line], 4.0 [dashed-dotted line] and 5.88 [dashed-(double)dotted
line]. We have used $G_\pi=13.51$\gev{-2}. (See Ref. [8].)}
\end{figure}

\begin{figure}
\includegraphics[bb=0 30 300 210, angle=0, scale=1]{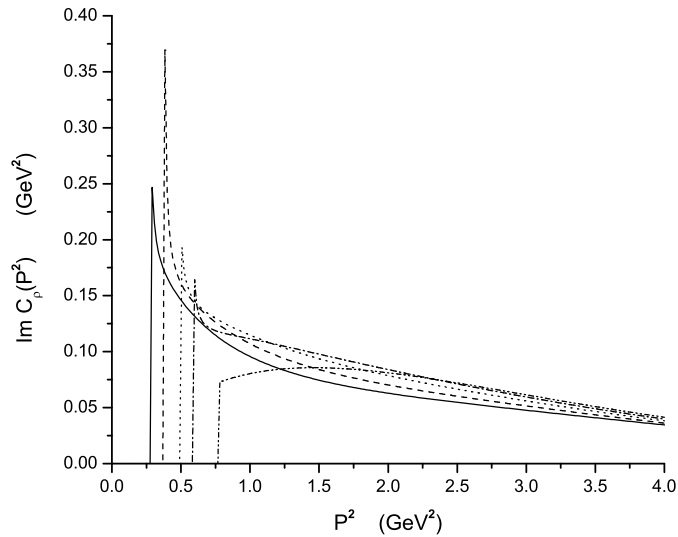}%
\caption{The figure shown the values of $\mbox{Im}C_\rho(P^2)$.
[See the caption of Fig. 14.] Here we have used
$G_V=11.46$\gev{-2}. (See Ref. [8].)}
\end{figure}

\begin{figure}
\includegraphics[bb=0 270 300 450, angle=0, scale=1.2]{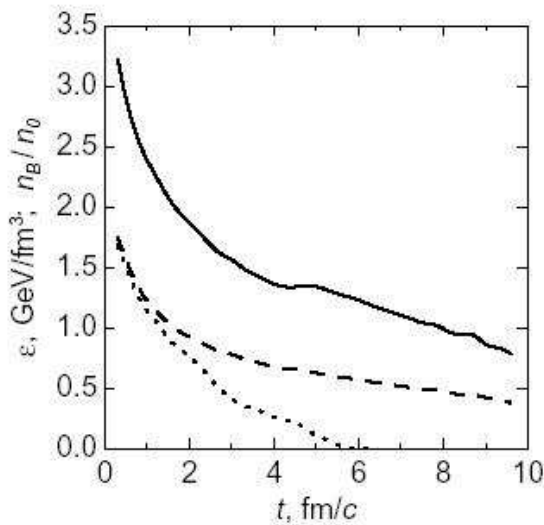}%
\caption{The average energy (solid line) and baryon (dashed)
densities of an expanding fireball formed in In+In collisions.
Dotted line shows a contribution of quarks and gluons to the
energy density, as calculated in Ref. [14].}
\end{figure}

\begin{figure}
\includegraphics[bb=0 265 400 410, angle=0, scale=1]{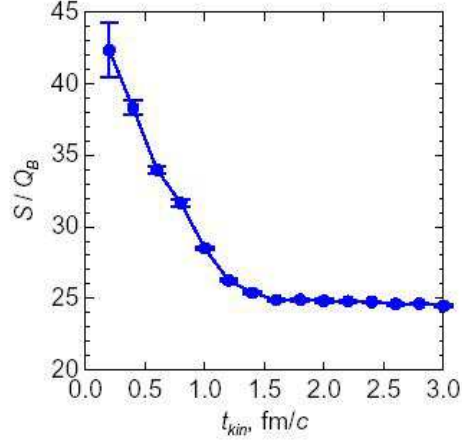}%
\caption{Temporal dependence of entropy $S$ per baryon charge
$Q_B$ of participants for semi-central In+In collision at
$E_{lab}=158A$ GeV, as calculated in Ref. [14].}
\end{figure}

\begin{figure}
\includegraphics[bb=0 245 400 450, angle=0, scale=1]{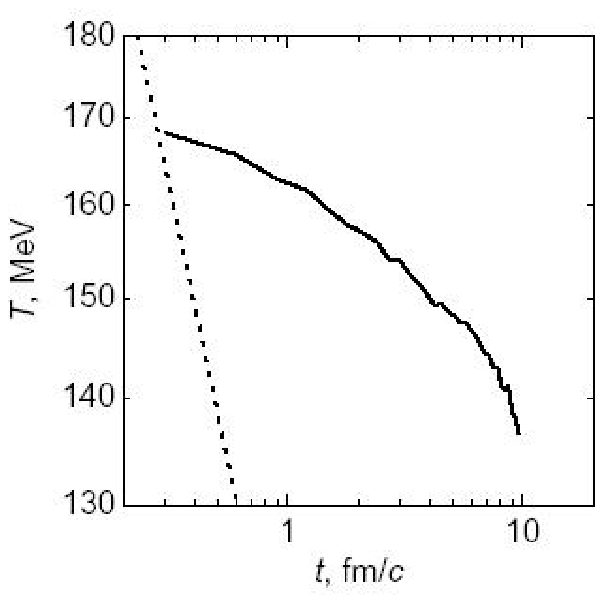}%
\caption{Evolution of the average temperature as calculated in
Ref. [14]. The dotted line corresponds to the Bjorken regime with
ultra-relativistic ideal gas EoS. (See Ref. [14].)}
\end{figure}

In Fig. 16, taken from Ref. [14], we show the calculated energy
density and values of $n_B/n_0=\rho/\rho_{NM}$ relevant to the
NA60 experiment. For times less than $t=1$ fm/c, $1.2\leq
n_B/n_0\leq1.7$.

In Fig. 17 the theoretical results for the ratio of the entropy
$S$ to the baryon charge in the NA60 experiment is shown, as
presented in Ref. [14] for the specific model used, the
Quark-Gluon String Model. The authors of Ref. [14] suggest that
for $t_{kin} \geq1.3$ fm/c the system may be considered as
undergoing isoentropic expansion.

In Fig. 18, taken from Ref. [14], the temperature is given as a
function of $t_{kin}$. For $0<t_{kin}<1$ fm/c the temperature is
in the range 162 MeV $\leq T\leq170$ MeV. While these values are a
bit below the deconfinement temperature at zero density, the value
of $n_B/n_0$ is given as 1.7 at $t=0$. The combination of the
elevated temperature and the finite matter density may be
sufficient to keep the system in the deconfined phase at the
earliest times of the collision, $t<1$ fm/c, as suggested in our
analysis.

The authors of Ref. [14] suggest that the critical temperature (at
the finite chemical potential $\mu_B$) is about 160 MeV. According
to Fig. 18, taken from Ref. [14], the temperature drops below 160
MeV for $t\geq1.3$ fm/c suggesting that the hadronic mode becomes
dominant above that temperature. According to Fig. 16, the baryon
density $n_B/n_0$ is approximately $1.0$ at $t\sim1.3$ fm/c. Note
that the dotted line in Fig. 16 represents the contribution of the
quarks and gluons to the energy density in the model of Ref. [14].

\section{Excitations of the Quark-Gluon Plasma at finite Temperature}

In this section we consider temperatures greater than $T_c$ and
present the hadronic correlation functions calculated using the
formalism of Ref. [7]. (We will not attempt to review that
formalism here, but only present some our results.) For example,
in Fig. 19 we show the correlation function in the
vector-isovector channel. The position of the peaks may be moved
by making small modifications of the coupling constant $G_V(T)$,
whose temperature dependence is not well known. For the coupling
constant that we have used for $T/T_c=1.2$, we find a peak in the
spectral function at about 600 MeV. (Here the value used for $G_V$
is equal to 0.8 times the value of $G_V$ for $T=0$. See the
Appendices for the definition of $G_V$.)

 \begin{figure}
 \includegraphics[bb=0 0 280 235, angle=0, scale=1]{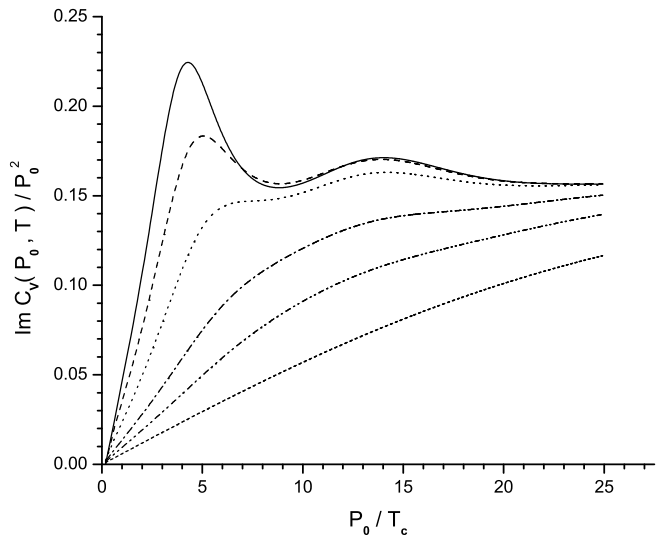}%
 \caption{Values of $\mbox{Im}\,C_V(P_0, T)/P_0^2$, obtained in Ref.\,[7], are shown for values of
 $T/T_c=1.2$ [\,solid line\,], $T/T_c=1.5$ [\,dashed line\,], $T/T_c=2.0$ [\,dotted line\,], $T/T_c=3.0$ [\,dot-dashed
 line\,], $T/T_c=4.0$ [\,double dot-dashed line\,], and $T/T_c=5.88$ [\,short dashed line\,]. Here
 we use $G_V(T)=G_V\,[1-0.17\,(T/T_c)]$ with $T_c=0.150$ GeV and $G_V=11.46\,\mbox{GeV}^{-2}$.}
 \end{figure}

\section{discussion}

We may return to a consideration of Fig. 1. We have suggested that
the large peak at about 750 MeV represents the observation of the
``prompt" leptons which are emitted for $t\leq 1$ fm/c when the
system is in a deconfined mode. (We have argued that the elevated
temperature and density is sufficient to deconfine the system at
the earliest stage of the collision.) As the system moves into the
confined phase for $t\geq1$ fm/c we see that the curve for
$S/Q_B$, seen in Fig. 17, changes its character, becoming constant
for $t>1.5$ fm/c. That is suggestive of the formation of a
confined phase in which we may discuss the validity of Brown-Rho
scaling. (We remark that the peaks seen in Figs. 14, 15 and 16 do
not represent bound-state mesons. Such mesons are deconfined at
the elevated temperature and densities.)

We suggest that in the confined phase, the system generates the
secondary peak seen in Fig. 1 at about 0.4-0.5 GeV. That is
roughly in accord with the dashed curve representing Eq. (1.3). We
also may suggest that the dashed curve peaks at a somewhat too low
an energy since Eq. (1.3) leads to $m_\rho^*=0$ at $T=T_c$ which
we believe overemphasizes the effect of temperature. (See Figs.
11-13 of Section III.)

If the interpretation of the NA60 data given in this work is
correct, we may argue that the measurement of lepton pairs from
vector-isovector states (resonances or bound states) can give us a
detailed picture of the evolution of the deconfined system to the
confined mode.

Recent work by Ruppert, Renk and M$\ddot{\mbox{{u}}}$ller contains
a discussion of the width of the rho meson in a nuclear medium
using QCD sum rules [26]. They are particularly concerned with how
the width of the rho mass in the medium will affect the Brown-Rho
scaling law \be
m_\rho^*/m_\rho\sim(<\bar{q}q>^*/<\bar{q}q>)^{1/2},\ee which is
the more recent form of the scaling law [27,28] than that given in
Ref. [1].

The widths of the rho in matter were calculated and presented in
Fig. 2 of Ref. [26]. For the parameter set corresponding to the
work of Hatsuda and Lee, the predicted width is approximately 300
MeV at $\rho/\rho_{NM}=1.5$. That is close to the width we may
read from Fig. 1, when we consider the first peak in that figure
at about 450-500 MeV. Therefore, our interpretation of the data is
not incompatible with the analysis of Ref. [26]. Additional
studies of the rho meson in matter may be found in Ref. [29].
Finally, we note that Brown and Rho have recently discussed the
NA60 data and the validity BR scaling in Refs. [30, 31].

A recent experiment which describes the in-medium modification of
the $\omega$ meson [37] provides support for BR scaling and the
argument put forth in the present work. Reference [37] reports
upon the photoproduction of the $\omega$ mesons on nuclei. They
result for the $\omega$ mass in matter may be put into the form
$m_\omega^*=m_\omega(1-0.14\rho/\rho_0)$ where $\rho_0$ is the
density of nuclear matter in the notation of Ref. [37]. We remark
that since the experiment does not involve the creation of high
temperature matter and a quark-gluon plasma, one does not expect
to see the large peak seen in Fig. 1 which we have ascribed to
excitations of the quark-gluon plasma with the quantum numbers of
the rho meson.

\appendix
\renewcommand{\theequation}{A\arabic{equation}}
  \setcounter{equation}{0}  
  \section{A model of confinement}

There are several models of confinement in use. One approach is
particularly suited to Euclidean-space calculations of hadron
properties. In that case one constructs a model of the quark
propagator by solving the Schwinger-Dyson equation. By appropriate
choice of the interaction one can construct a propagator that has
no on-mass-shell poles when the propagator is continued into
Minkowski space. Such calculations have recently been reviewed by
Roberts and Schmidt [32]. In the past, we have performed
calculations of the quark and gluon propagators in Euclidean space
and in Minkowski space. These calculations give rise to
propagators which did not have on-mass-shell poles [33-36].
However, for our studies of meson spectra, which included a
description of radial excitations, we found it useful to work in
Minkowski space.

The construction of our covariant confinement model has been
described in a number of works. In all our work we have made use
of Lorentz-vector confinement, so that the Lagrangian of our model
exhibits chiral symmetry. We begin with the form $V^C(r)=\kappa
r\mbox{exp}[-\mu r]$ and obtain the momentum-space potential via
Fourier transformation. Thus, \be V^C(\vec k-\vec
k\,^\prime)=-8\pi\kappa\left[\frac1{[(\vec k-\vec
k\,^\prime)^2+\mu^2]^2}-\frac{4\mu^2}{[(\vec k-\vec
k\,^\prime)^2+\mu^2]^3}\right]\,,\ee with the matrix form \be
\overline V{}\,^C(\vec k-\vec k\,^\prime)=\gamma^\mu(1)V^C(\vec
k-\vec k\,^\prime)\gamma_\mu(2)\,,\ee appropriate to
Lorentz-vector confinement. The potential of Eq. (A1) is used in
the meson rest frame. We may write a covariant version of
$V^C(\vec k-\vec k^\prime)$ by introducing the four-vectors \be
\hat k^\mu=k^\mu-\frac{(k\cdot P)P^\mu}{P^2}\,, \ee  and \be \hat
k^{\prime\,\mu}=k^{\prime\,\mu}-\frac{(k^\prime\cdot
P)P^\mu}{P^2}\,. \ee Thus, we have \be V^C(\hat k-\hat
k\,^\prime)=-8\pi\kappa\left[\frac1{[-(\hat k-\hat
k\,^\prime)^2+\mu^2]^2}-\frac{4\mu^2}{[-(\hat k-\hat
k\,^\prime)^2+\mu^2]^3}\right]\,.\ee Originally, the parameter
$\mu=0.010$ GeV was introduced to simplify our momentum-space
calculations. However, in the light of the following discussion,
we can remark that $\mu$ may be interpreted as describing
screening effects as they affect the confining potential [25]. In
our work, we found that the use of $\kappa=0.055$\gev2 gave very
good results for meson spectra.

The potential $V^C(r)=\kappa r\mbox{exp}[-\mu r]$ has a maximum at
$r=1/\mu$, at which point the value is $V_{max}=\kappa/\mu e=
2.023$ GeV. If we consider pseudoscalar mesons, which have $L=0$,
the continuum of the model starts at $E_{cont}=m_1+m_2+V_{max}$,
so that for $m_1=m_2=m_u=m_d=0.364$ GeV, $E_{cont}=2.751$ GeV. It
is also worth noting that the potential goes to zero for very
large r. Thus, there are scattering states whose lowest energy
would be $m_1+m_2$. However, barrier penetration plays no role in
our work. The bound states in the interior of the potential do not
communicate with these scattering states to any significant
degree. It is not difficult to construct a computer program that
picks out the bound states from all the states found upon
diagonalizing the random-phase-approximation Hamiltonian.

\renewcommand{\theequation}{B\arabic{equation}}
  \setcounter{equation}{0}  
  \section{density and temperature dependence of the confining field}

In part, our study of the confining interaction has been
stimulated by the results presented in Ref. [25] for the
temperature-dependent potential, $V(r)$, in the case dynamical
quarks are present. We reproduce some of the results of that work
in Fig. 20. There, the filled symbols represent the results for
$T/T_c=0.68, 0.80, 0.88$ and 0.94 when dynamical quarks are
present. This figure represents definite evidence of ``string
breaking", since the force between the quarks appears to approach
zero for $r > 1$ fm. This is not evidence for deconfinement, which
is found for $T=T_c$. Rather, it represents the creation of a
second $\bar qq$ pair, so that one has two mesons after string
breaking. Some clear evidence for string breaking at zero
temperature and finite density is reported in Ref. [25].

 \begin{figure}
 \includegraphics[bb=0 0 250 350, angle=-90, scale=1]{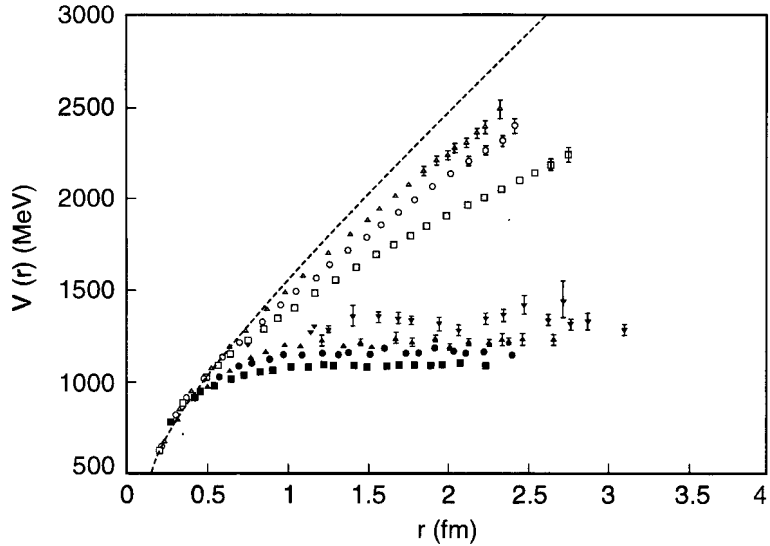}%
 \caption{A comparison of quenched (open symbols) and unquenched results (filled symbols) for
 the interquark potential at finite temperature [25]. The dotted line is the zero temperature
 quenched potential. Here, the symbols for $T=0.80T_c$ [open triangle], $T=0.88T_c$
 [open circle], $T=0.80T_c$ [open square], represent the quenched
 results. The results with dynamical fermions are given at $T=0.68T_c$ [solid downward-pointing
 triangle], $T=0.80T_c$ [solid upward-pointing triangle], $T=0.88T_c$ [solid circle],
 and $T=0.94T_c$ [solid square].}
 \end{figure}

In order to study deconfinement in our generalized NJL model, we
need to specify the interquark potential at finite density. In
that case we had used $V^C(r)=\kappa r\mbox{exp}[-\mu r]$ for zero
matter density. For the model we study in this work, we write \be
V^C(r, \rho)=\kappa r\mbox{exp}[-\mu(\rho) r]\ee and put \be
\mu(\rho)=\frac{\mu_0}{1-\left(\displaystyle\frac\rho{\rho_C}\right)^2}\,,\ee
with $\rho_C=2.25\rho_{NM}$ and $\mu_0=0.010$ GeV. With this
modification our results for meson spectra in the vacuum are
unchanged. Other forms than that given in Eqs. (B1) and (B2) may
be used. However, in this work we limit our analysis to the model
described by these equations. The corresponding potentials for our
model of Lorentz-vector confinement are shown in Fig. 21 for
several values of $\rho/\rho_{NM}$. In the case of finite
temperature we make use of the potentials $V^c(r,T)$ shown in Fig.
22.

 \begin{figure}
 \includegraphics[bb=0 20 250 220, angle=0, scale=1]{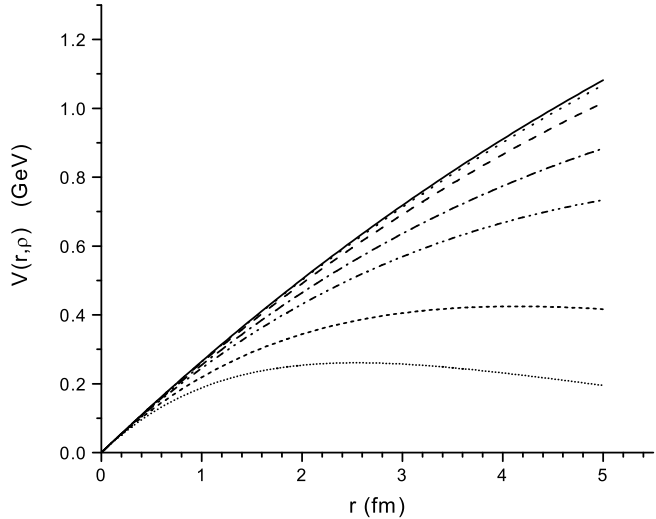}%
 \caption{Values of $V(r,\rho)$  are shown, where $V(r,\rho)=\kappa r\exp[-\mu(\rho)r]$
 and $\mu(\rho)=\mu_0/[1-(\rho/\rho_C)^2]$. Here $\rho_C=2.25\rho_{NM}$ and $\mu_0=0.010$ GeV.
 The values of $\rho/\rho_{NM}$ are 0.0 [solid line], 0.50 [dotted line], 1.0 [dashed line],
 1.50 [dashed-dotted line]. 1.75 [dashed-dotted-dotted line], 2.0 [short-dashed line],
 and 2.1 [small dotted line].}
 \end{figure}

 \begin{figure}
 \includegraphics[bb=0 0 300 200, angle=0, scale=1]{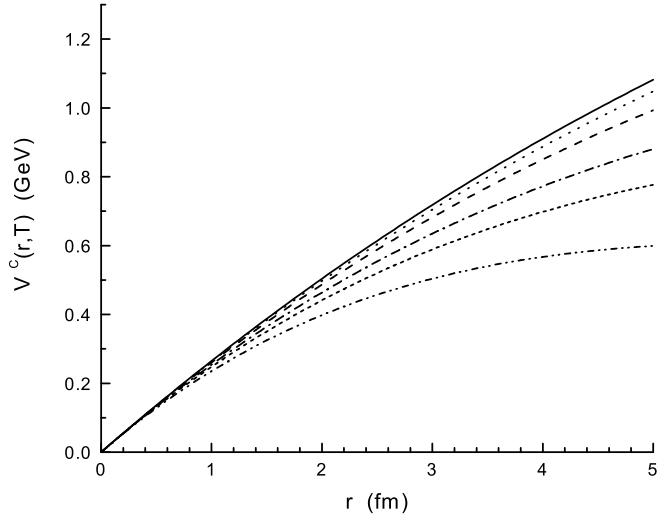}%
 \caption{The potential $V^C(r, T)$ is shown for $T/T_c=0$ [solid line],
 $T/T_c=0.4$ [dotted line], $T/T_c=0.6$ [dashed line], $T/T_c=0.8$ [dashed-dotted line],
 $T/T_c=0.9$ [short dashes], $T/T_c=1.0$ [dashed-(double) dotted line]. Here,
 $V^C(r,T)=\kappa r\exp[-\mu(T)r]$, with $\mu(T)=0.01\mbox{GeV}/[1-0.7(T/T_c)^2]$ and
 $\kappa=0.055$\gev2.}
 \end{figure}



\end{document}